# Synthetic chiral magnets promoted by the Dzyaloshinskii–Moriya interaction

Aleš Hrabec[1, 2, 3], Zhaochu Luo[1, 2], Laura J. Heyderman[1, 2], and Pietro Gambardella[3]

[1]Laboratory for Mesoscopic Systems, Department of Materials, ETH Zurich, 8093 Zurich, Switzerland

[2]Paul Scherrer Institut, 5232 Villigen PSI, Switzerland

[3]Laboratory for Magnetism and Interface Physics, Department of Materials, ETH Zurich, 8093 Zurich, Switzerland

Authors to whom correspondence should be addressed: ales.hrabec@psi.ch; laura.heyderman@psi.ch; pietro.gambardella@mat.ethz.ch

**Abstract**

The ability to engineer the interactions in assemblies of nanoscale magnets is central to the development of artificial spin systems and spintronic technologies. Following the emergence of the Dzyaloshinskii-Moriya interaction (DMI) in thin film magnetism, new routes have been opened to couple the nanomagnets via strong chiral interactions, which is complementary to the established dipolar and exchange coupling mechanisms. In this Perspective, we review recent progress in the engineering of synthetic magnets coupled by the interlayer and intralayer DMI. We show how multilayer chiral magnetic structures and two-dimensional synthetic antiferromagnets, skyrmions, and artificial spin systems can be realized by simultaneous control of the DMI and magnetic anisotropy. In addition, we show that, with the combination of DMI and current-induced spin-orbit torques, field-free switching of synthetic magnetic elements is obtained as well as all-electric domain wall logic circuits.

**Main Text**

Synthetic magnets are assemblies of coupled magnetic elements whose dimensions, position, and properties can be tuned to fix the relative orientation of their magnetization and determine their response to external stimuli, such as magnetic fields and electric currents. A Halbach array, initially developed to focus particle beams[1], is an example of a macroscopic assembly of permanent magnets interacting via the dipole interaction, that augments and confines the magnetic stray field outside the array. At the microscopic scale, synthetic arrays of magnets with dimensions ranging from micrometers down to a few tens of nanometers can be designed to obtain magnetic configurations with a well-defined hierarchy of energy levels and degeneracies, that is determined by the competing dipolar interactions among the elements of the array[2–4]. These microscopic arrays are extensively investigated for a wide range of uses including nanomagnetic logic gates[5,6], magnetic metamaterials[7,8], magnetic micromachines[9] and spin ices[10–14]. The dipolar-coupled synthetic magnets that make up these arrays include both vertically-stacked and coplanar structures, as schematically shown in Figs. 1a and b, respectively.

As dipolar interactions decrease when reducing the volume of the magnetic elements, it is not practical to realize ultrathin synthetic magnets based on these interactions for applications in electronics and spintronics. Instead, short-ranged quantum mechanical interactions provide suitable alternatives to couple magnetic films in multilayers and nanostructures. These include the exchange bias between a ferromagnetic and an antiferromagnetic layer[15] (Fig. 1c) and the oscillatory Ruderman-Kittel-Kasuya-Yosida (RKKY) interaction between two magnetic layers separated by a nonmagnetic metal[16] (Fig. 1d). Synthetic antiferromagnets consisting of RKKY-coupled ferromagnetic layers, for example, are widely employed to tune the magnetic hysteresis of spin valves and magnetic tunnel junctions used in sensing and data storage applications[17–19]. The possibility to design synthetic magnets for specific applications is thus intimately related to the ability to induce and tune the couplings between the individual magnetic elements.

With the emergence of the Dzyaloshinskii–Moriya interaction (DMI)[20,21] as a tool to engineer the magnetic texture of thin films with asymmetric interfaces[22–24], fascinating prospects are now open for realizing novel types of synthetic magnets, which are the subject of this Perspective. The DMI arises in magnetic systems that lack a centre of inversion and exhibit strong spin–orbit coupling. This can occur in bulk crystalline materials[25], amorphous alloys with a composition gradient[26], as well as in thin magnetic films where the conditions for the emergence of the DMI are naturally fulfilled at asymmetric interfaces, such as for ferromagnets in contact with heavy metals[27–30] or graphene[31,32], and ferrimagnetic garnets grown on nonmagnetic substrates[33,34]. This interaction can be represented by the Hamiltonian $H_{DMI} = -\Sigma_{ij} \boldsymbol{D}_{ij} \cdot (\boldsymbol{S}_i \times \boldsymbol{S}_j)$, where two neighbouring spins $\boldsymbol{S}_i$ and $\boldsymbol{S}_j$ are indirectly coupled via a heavy metal atom. The DMI thus favours an orthogonal alignment between adjacent magnetic moments with the sense of rotation of the magnetic moments in space determined by the direction of the vector $\boldsymbol{D}$. Therefore, in stark contrast to the dipolar and exchange interactions, the DMI is chiral and the resulting magnetic textures are chiral too. Depending on the magnitude of $\boldsymbol{D}$ relative to the exchange interaction and magnetic anisotropy, the DMI can induce the formation of chiral Néel domain walls[33,35–40] or noncollinear magnetic ground states, such as spin helices, cycloids, or skyrmions[41–49].

The DMI favours the noncollinear alignment of neighboring spins, thus inducing a twist in the magnetic texture. The extension of such a twist in systems with spatially homogeneous magnetic properties is given to a first approximation by the competition between the exchange interaction, DMI and magnetic anisotropy[50] and is typically on the order of 10 nanometers[44,51,52] Conversely, magnetic systems that are designed with non-homogeneous properties, such as magnetic multilayers and planar arrays of nanomagnets, provide an ideal playground to investigate the DMI-mediated chiral coupling for a wider range of length scales and geometries. In such structures, the magnetization of neighbouring magnetic regions follows the specific chirality inherited from the DMI. This magnetic chirality can take place either in vertical stacks via the so-called interlayer DMI coupling[53–57] (Fig. 1e) or in planar non-homogeneous structures, where the magnetic textures are connected via intralayer lateral chiral coupling[58–60] (Fig. 1f).

The *interlayer* DMI coupling exploits the antisymmetric part of the RKKY interaction[22,56]. Early reports provided evidence of this type of coupling in Dy/Y magnetic multilayers, which possess a helical spin structure with a predominant chirality induced by an in-plane magnetic field[53,54]. More recently, interlayer chiral coupling has also been demonstrated in transition metal systems. For example, a CoFeB layer with in-plane (IP) magnetic anisotropy coupled via a PtRuPt spacer to a thin Co layer with out-of-plane (OOP) anisotropy displays chirally-canted magnetic configurations due to the interplay of antiferromagnetic RKKY coupling and DMI[55]. Similarly, ferromagnetically (Fig. 2a) and antiferromagnetically (Fig. 2b) coupled synthetic magnets consisting of two Co layers separated by a PtRuPt spacer have asymmetric OOP hysteresis loops in the presence of a static in-plane magnetic field, which is consistent with the chirality of the DMI[57]. Although the interlayer DMI is weaker than the collinear RKKY interaction in these systems, further tuning of the composition and thickness of the nonmagnetic spacer might lead to larger net interlayer DMI energies[56]. This, in turn, would enable the realization of three-dimensional topological multilayer structures. Moreover, functionalities might be integrated in synthetic magnets, such as asymmetric effects in the propagation of domain walls and spin waves[61].

The *intralayer* DMI is typically stronger than the *interlayer* DMI. Whereas the first gives a surface energy contribution of the order of 1-2 mJ m$^{-2}$ in metallic ferromagnets in contact with heavy metals such as Co/Pt[60,62], the second is at least one order of magnitude smaller owing to the presence of a nonmagnetic spacer layer between the two ferromagnetic layers[55–57]. In combination with high-resolution patterning and nanofabrication techniques, the lateral chiral coupling promoted by the *intralayer* DMI can lead to unprecedented control of the magnetic interactions in planar arrays of nanomagnets. This has been recently demonstrated for different arrangements of nanomagnets with IP and OOP magnetization patterned out of a continuous Pt/Co/Al trilayer[60], as shown in Fig. 3. Here, the different regions with IP and OOP anisotropy are defined by selective oxidation of the Al capping layer[63]. Whereas the anisotropy of the as-grown films has been adjusted to be IP, the contribution to the interfacial anisotropy from the Co/AlOx interface can generate an OOP magnetization. Additionally, the Pt underlayer connects the regions with different anisotropies via the DMI, which leads to a left-handed chiral alignment of the magnetic moments in the neighboring IP-OOP regions, such that the left-handed →⊙ and ←⊗ orientations are energetically favoured over the right-handed ⊙→ and ⊗← orientations (Fig. 1f). It has been shown that the strength of the chiral coupling can be significantly larger than the energy barrier for magnetization reversal between ⊙ and ⊗ states or → and ← states. Thus, chiral ordering spontaneously appears in elongated nanomagnets composed of one or more sequences of IP-OOP elements (Figs 3a, b and c). Moreover, switching of the IP (OOP) element with an external IP (OOP) magnetic field automatically leads to the switching of the OOP (IP) element attached to it. The energetic imbalance between IP-OOP magnetic orientations with opposite chirality also leads to the appearance of a lateral exchange bias, as seen in Fig. 4a, and to the appearance of stable and metastable multistate magnetic configurations[60].

The lateral chiral coupling can be tuned by material and interface engineering as well as by changing the size of the nanomagnets, since the DMI is proportional to the length of the quasi one-dimensional domain wall separating the OOP and IP elements. The strength of the coupling will in turn define the obtainable magnetic configurations as demonstrated by the lateral synthetic antiferromagnets in Figs. 3b and c. As long as the IP spacer is shorter than about 100 nm and retains a monodomain state, the two OOP elements effectively interact with each other. Thus, reversing the magnetization of one OOP element with an OOP field causes the reversal of the IP spacer and subsequently the reversal of the other OOP element. Furthermore, such coupled OOP and IP elements can be used as building blocks of more complex spin systems with a well-defined symmetry and topology. For example, synthetic Néel skyrmions with an arbitrary number of windings can be realized by coupling concentric OOP rings with 50 nm wide IP rings (Fig. 3d). In addition, it is possible to design mesoscopic two-dimensional structures such as artificial spin ice and other artificial spin systems with either frustrated or non-frustrated interactions[60] as shown in Figs. 3e and f, respectively. Since the chiral coupling in thin magnetic films is more efficient than the dipolar coupling, such chirally-coupled arrays of nanomagnets are energetically robust up to room temperature and beyond. They can also shed more light on the ordering of complex systems[64,65], as well as on the transport and magneto-thermal effects in non-collinear magnetic structures[66–68].

In planar systems, the heavy metal underlayer offers more than a platform for creating chirally coupled structures. If an electric current is passed through the Pt, spin-orbit coupling promotes the scattering and polarization of spins, as in the spin Hall and Rashba-Edelstein effects, leading to a spin accumulation at the interface with an adjacent ferromagnet[69]. The spin-orbit torque resulting from the diffusion and absorption of the spin current inside the ferromagnet can be used to switch the magnetization of both OOP and IP layers[69–74]. However, OOP magnetized elements generally require a static in-plane magnetic field to break the torque symmetry and achieve deterministic switching[69]. For chirally coupled OOP-IP nanomagnets, the built-in symmetry-breaking mechanism, means that the OOP (and therefore the coupled IP) magnetization can be switched by an electric current in the absence of external fields (Fig. 4a). Since the state of such a magnetic element can be read out via the anomalous Hall effect (Fig. 4a), chirally-coupled magnets constitute a basic unit for data storage.

The chiral coupling can also be incorporated into magnetic tracks where the spin-orbit torques can be further exploited for several different applications which can be directly implemented in racetrack memories[75]. Controllable and efficient domain wall injection and deletion[76–78] is a pre-requisite for memories relying on domain wall motion. One possibility is to insert coupled IP-OOP boundaries as domain wall injectors (Fig.4b), where the switching of the magnetization of the IP region within a racetrack by an external magnetic field can enable or disable the injection of a magnetic domain wall in an OOP track. This has been realized by patterning nanowires comprising of a short IP region at the beginning of the track followed by a longer OOP region, with width ranging from 100 nm to 4 μm and length up to several μm[79]. The switching of the IP part was achieved by an external magnetic field of a few mT, which in principle can be replaced also by a local spin-orbit torque. In addition, magnetic tunnel junctions patterned on top of the racetrack can be used to control the orientation of the IP region or directly inject domain walls in the OOP region. The patterning of IP regions inside OOP tracks can further be used to pin domain walls at specific locations[80] and tune the domain wall chirality[81].

The current-induced switching and propagation of magnetic domains combined with the multiplicity of magnetic states that can be stabilized in chiral coupled OOP-IP elements opens entirely new opportunities for the realization of all-electric nanomagnetic logic circuits[82]. The key point here is that lateral chiral coupling provides a way to design magnetic domain wall inverters and majority logic gates that are otherwise difficult to implement in racetrack structures[5,6]. When a narrow IP region is incorporated into an OOP magnetized track, it couples to its surrounding, leading to the antiferromagnetic alignment of the OOP magnetization on the left and right of the IP region.

When a domain wall propagating in the track encounters such a region and the current-induced spin-orbit torque driving the domain wall is large enough, the magnetization of the IP regions flips, leading to the annihilation of the incoming domain wall and the nucleation of a new domain wall of opposite polarity on the other side of the IP region (Fig. 5a,b). The chiral OOP-IP-OOP structure therefore serves as an inverter capable of transforming an up-down (down-up) domain wall into a down-up (up-down) domain wall[82]. If the down and up magnetization directions in the racetrack represent the Boolean logic "1" and "0", respectively, this process is equivalent to the NOT logical operation. Hence, the magnetic track not only serves as a data storage device but can also perform logic operations. Based on this principle, one can design a majority gate where three OOP tracks (two inputs and one bias) meet at a common point delimited by an IP region that is connected to an output track (see Fig. 5c). Depending on the magnetization direction of the OOP input tracks next to the IP boundary, the output magnetization flips according to the truth table of either a NAND or a NOR gate, where the logic function is determined by the magnetization of the bias terminal[82]. As these gates are functionally complete and can be cascaded on the same current line, in principle any type of logic operation can be performed, as exemplified by the full adder gate shown in Fig. 5d. Chirally coupled logic gates thus provide a means to integrate memory and logic functions in a single platform. These gates are scalable, operated by electric currents, and are compatible with back-end-of-line processes established for the fabrication of magnetic random access memories.

In order to optimize chiral synthetic magnets for specific applications, key design requirements must be met, including the ability to tune the DMI and/or the magnetic anisotropy of the individual magnetic elements with high spatial accuracy. In magnetic multilayers, this can be achieved by changing the thickness of the nonmagnetic spacers and engineering the anisotropy of the individual layers in the stack. In planar structures, the DMI and interfacial magnetic anisotropy are often not independent of each other. Starting from a system with strong DMI, the most effective approach to design such a synthetic magnet relies on the patterning of the magnetic anisotropy. This patterning can be achieved in a controlled way by spatially varying the degree of oxidation of the magnetic interfaces[83,84]. Similar effects can be obtained with focused-ion beam irradiation using Ga or He ions[38,85–87], local thermal annealing[88], and electric gating[89–94]. Whereas modification with a focused ion beam has the strongest miniaturization potential, with a patterning resolution down to 5 nm[95], electric gating would provide a method to switch the chiral coupling on and off[96]. Such techniques might be employed to manufacture electrically reconfigurable logic gates, unconventional computation building blocks[97,98] or magnonic devices[88,99].

In magnetic thin films lacking an intrinsic DMI, synthetic chiral structures can be obtained by patterning curvilinear shapes[100] and defects[101], in which the gradient of the local curvature induces an effective chiral magnetic field originating from the exchange interaction[102,103]. In addition, three-dimensional chiral structures can be realized by geometric design[104], as is the case for magnetic helices and double-helices, which can host nontrivial topological textures[105] and present unusual optical[106] magnetoelectric[107] and magnetoresistive properties[108]. These topics provide interesting avenues of research in the field of synthetic chiral magnets.

In conclusion, the multiplicity of magnetic states that can be stabilized in interlayer and intralayer chirally-coupled structures opens the way to design synthetic magnets with great flexibility, scalability, and tunable collective behavior that is complementary to other long-established methods. The possibility to combine current-driven magnetic switching with the propagation of domain walls in such structures means that novel spintronic devices can be realised, such as nanomagnetic logic gates and neuro-inspired circuits. Furthermore, the topology and magnetic landscape of synthetic media can be reconfigured by electric currents or gating.

**Acknowledgments**

A.H. was funded by the European Union's Horizon 2020 research and innovation programme under Marie Sklodowska-Curie grant agreement no. 794207 (ASIQS).

**Data availability**

Data sharing is not applicable to this article as no new data were created in this study.

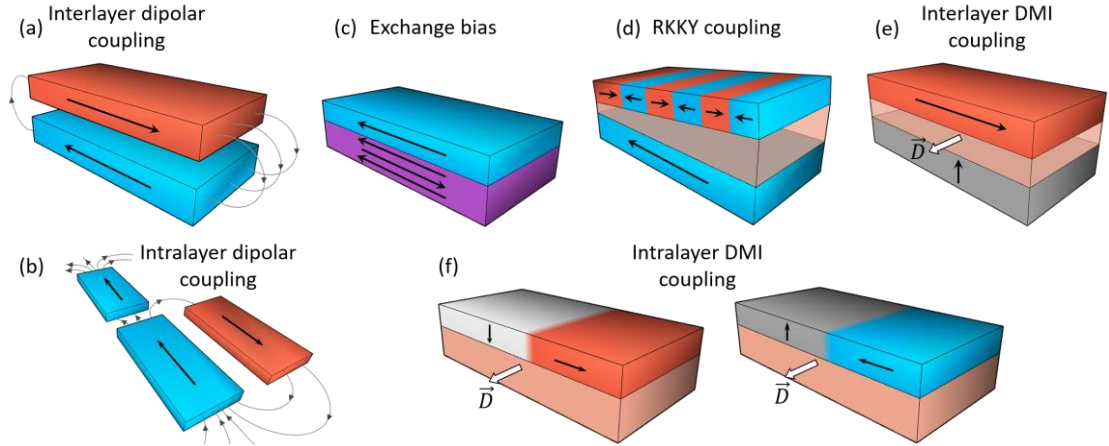

Figure 1. Synthetic magnets with various magnetic coupling mechanisms: (a) Interlayer and (b) intralayer dipolar coupling, (c) exchange bias, (d) RKKY coupling, (e) interlayer and (f) intralayer DMI coupling. The orientation of $\vec{D}$ displayed in (e) reflects the global effect of the DMI on the coupling between the top and bottom layer. The actual magnitude and orientation of $\vec{D}$ depends on the interfacial atom positions and crystallographic orientation.

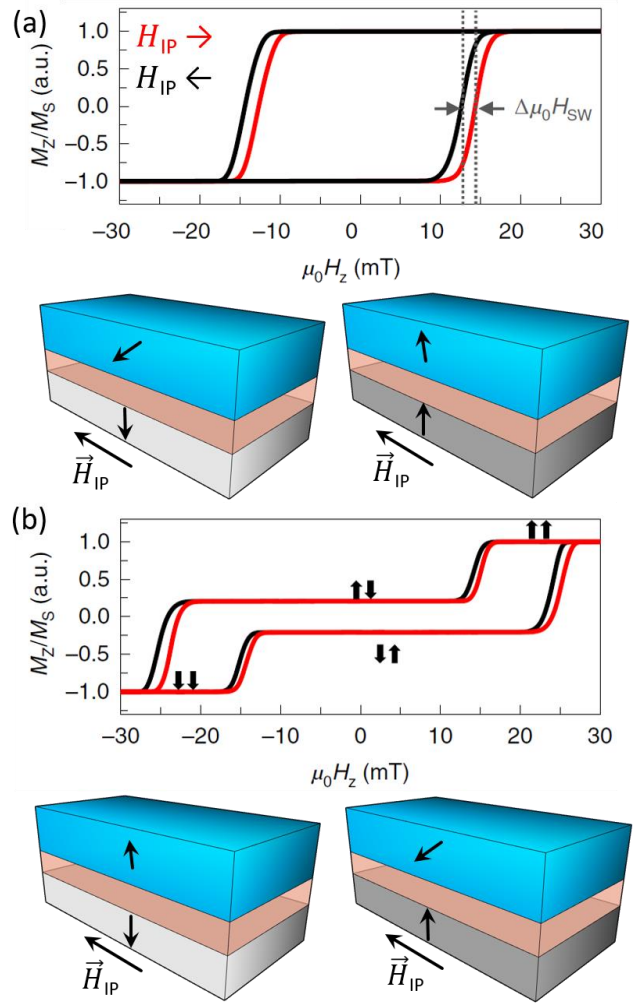

Figure 2. The interlayer DMI in an unbalanced synthetic magnet is shown to reduce or enhance the energy barrier to reverse the magnetization of the thicker layer with a lower anisotropy, as reflected by the shift of the hysteresis loops with respect to each other by $\Delta\mu_0 H_{sw}$. The magnetization tilts differently in (a) ferromagnetically and (b) antiferromagnetically coupled magnetic layers under application of a positive (red) and negative (black) in-plane magnetic fields. Reproduced from [57]

and reprinted with permission from Springer Nature.

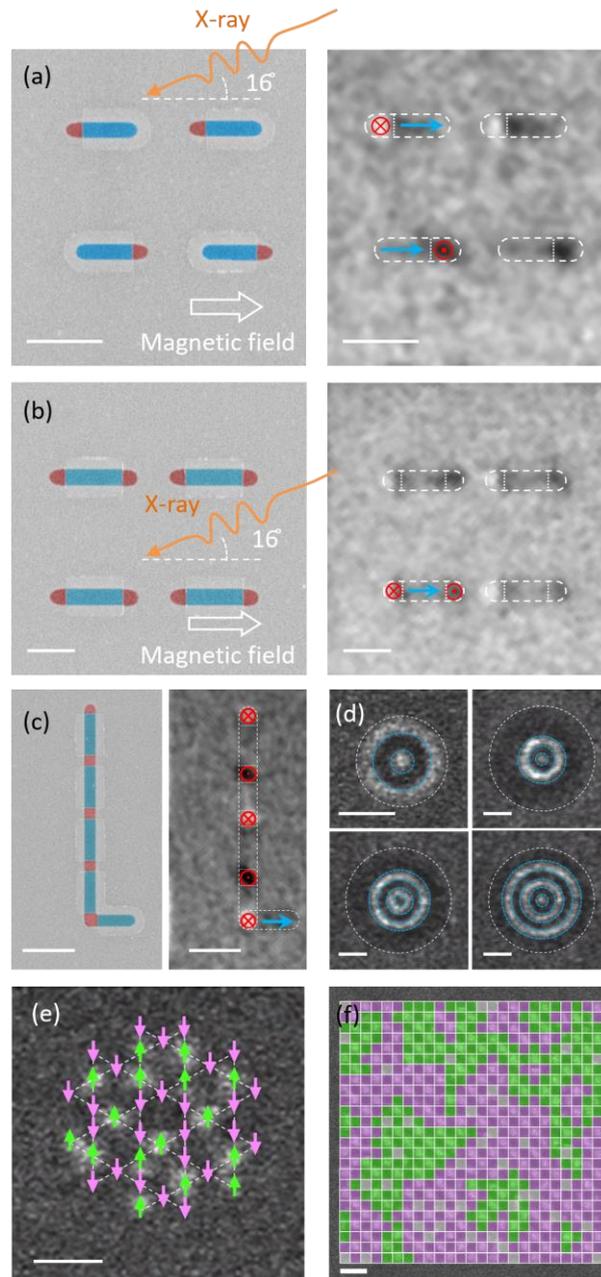

Figure 3. Synthetic chiral magnets in 2D structures: (a-c) Scanning electron micrographs [left panel] and x-ray photoemission electron micrographs [right panel] of (a) IP-OOP magnets, (b,c) OOP-IP-OOP lateral synthetic antiferromagnets. The dark and bright contrast in the out-of-plane regions corresponds to ⊙ and. ⊗ states, respectively. The blue and red coloring of scanning electron images corresponds to IP and OOP magnetized regions, respectively. (d) Magnetic force microscopy images of synthetic skyrmions. The IP 50 nm-wide magnetized parts are indicated by blue dashed lines; the boundary of the magnetic disk is indicated by white dashed line. (e,f) Realization of an artificial spin system arranged on (e) a kagome and (f) a square lattice. All scale bars are 500 nm. Reproduced from [60] and reprinted with permission from AAAS.

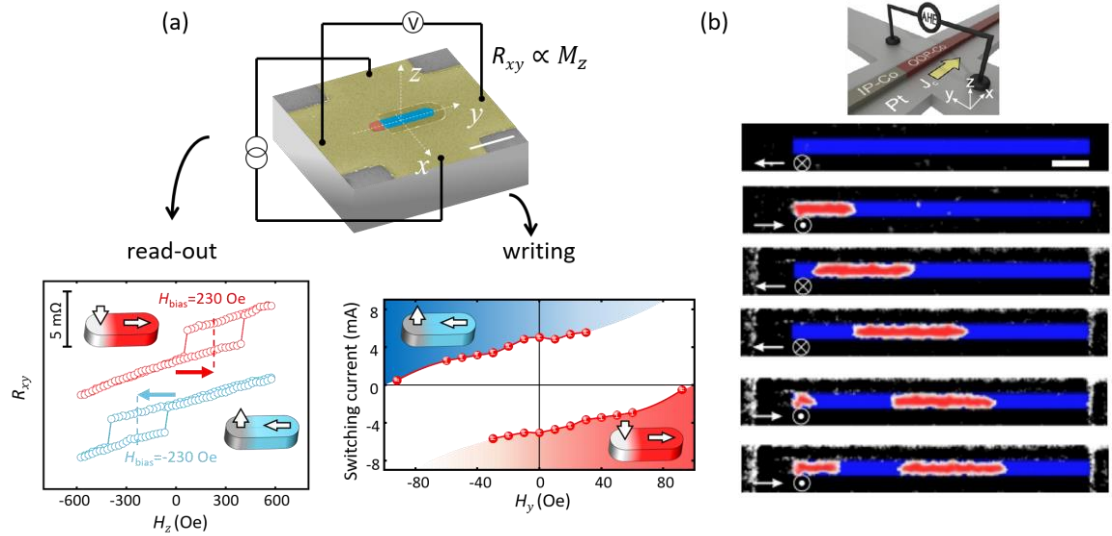

Figure 4. Different functionalities driven by an electric current in chirally coupled nanomagnets: (a) Schematic of an OOP-IP element fabricated on top of a Pt Hall cross allowing electrical read-out and writing of the magnetization. The readout of the nanomagnet is carried out via the anomalous Hall effect resistance $R_{xy}$, which reflects the z-component of the magnetization. Coupling to the IP magnet results in lateral exchange bias, as shown by the shifted magnetization loops. The sign of the bias is determined by the magnetization of the IP magnet. Field-free switching of the OOP magnetization in the OOP-IP element is achieved by injecting an electric current in the Pt Hall cross. Reproduced from [60] and reprinted with permission from AAAS. The scale bar corresponds to 2 µm. (b) current-driven chiral domain-wall injection into an OOP track. The scale bar corresponds to 2 µm. Reproduced from [79] and reprinted with permission from American Chemical Society.

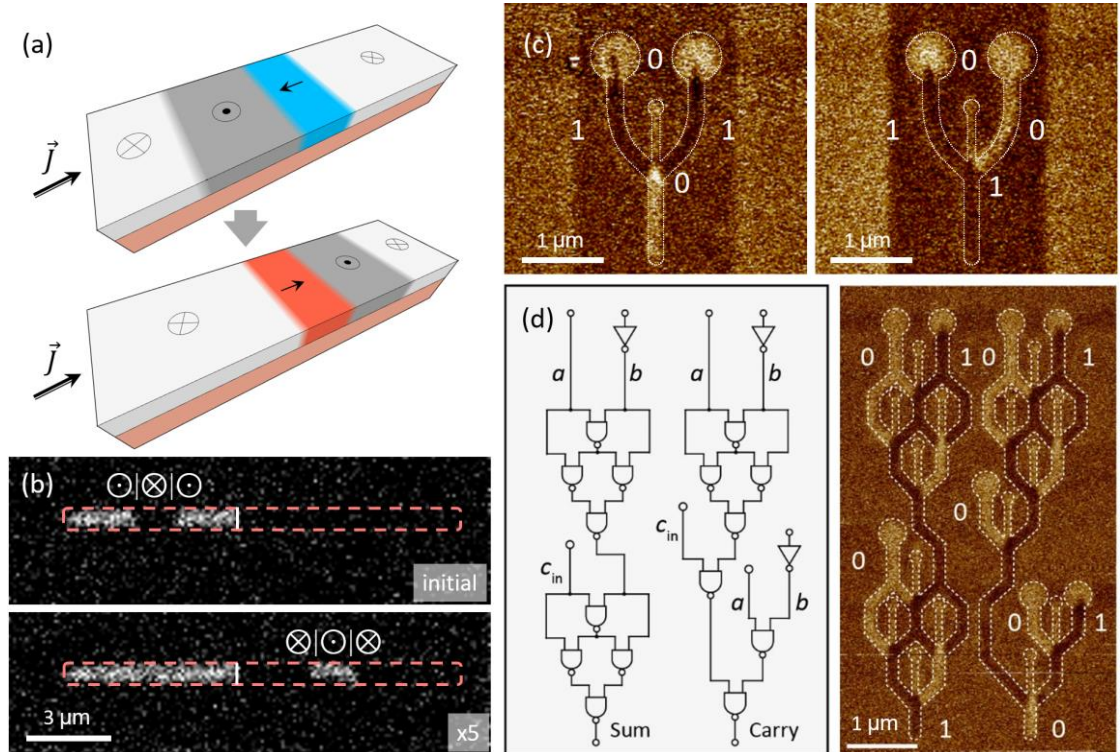

Figure 5. Current-driven magnetic domain-wall logic circuits. (a) Schematics showing the principle of current-driven domain-wall inversion (NOT gate). A ⊗|⊙ domain wall is inverted to give an ⊙|⊗ domain wall. (b) Magneto-optical Kerr effect images of domain-wall inverters before and after application of a series of electric pulses. An ⊙|⊗|⊙ domain is inverted to give a ⊗|⊙|⊗ domain. The inverter is indicated by a white line. (c) Magnetic force microscopy images of NAND

gates with inputs of "11" and "10" resulting in "0" and "1" outputs, respectively. (d) Left panel: Schematic of a full adder gate obtained by cascading several NAND gates. Right panel: Full adder gate with "$a$=0" and "$b$=1" inputs resulting in "Sum=1" with a "Carry = 0". Reproduced from [82] and reprinted with permission from Springer Nature.

[1] J.C. Mallinson, IEEE Trans. Magn. **9**, 678 (1973).
[2] R.P. Cowburn and M.E. Welland, Science **287**, 1466 (2000).
[3] C.A. Ross, S. Haratani, F.J. Castaño, Y. Hao, M. Hwang, M. Shima, J.Y. Cheng, B. Vögeli, M. Farhoud, M. Walsh, and H.I. Smith, J. Appl. Phys. **91**, 6848 (2002).
[4] J.I. Martín, J. Nogués, K. Liu, J.L. Vicent, and I.K. Schuller, J. Magn. Magn. Mater. **256**, 449 (2003).
[5] A. Imre, G. Csaba, L. Ji, A. Orlov, G.H. Bernstein, and W. Porod, Science **311**, 205 (2006).
[6] D.A. Allwood, G. Xiong, C.C. Faulkner, D. Atkinson, D. Petit, and R.P. Cowburn, Science **309**, 1688 (2005).
[7] M. Krawczyk and D. Grundler, J. Phys. Condens. Matter **26**, 123202 (2014).
[8] M. Decker, M.W. Klein, M. Wegener, and S. Linden, Opt. Lett. **32**, 856 (2007).
[9] J. Cui, T.Y. Huang, Z. Luo, P. Testa, H. Gu, X.Z. Chen, B.J. Nelson, and L.J. Heyderman, Nature **575**, 164 (2019).
[10] L.J. Heyderman and R.L. Stamps, J. Phys. Condens. Matter **25**, 363201 (2013).
[11] C. Nisoli, R. Moessner, and P. Schiffer, Rev. Mod. Phys. **85**, 1473 (2013).
[12] S.H. Skjærvø, C.H. Marrows, R.L. Stamps, and L.J. Heyderman, Nat. Rev. Phys. **2**, 13 (2020).
[13] L. Yu, Y. Wang, J. Li, F. Zhu, X. Meng, J. Cao, C. Jing, Y. Wu, and R. Tai, AIP Adv. **7**, 085211 (2017).
[14] J. Lehmann, C. Donnelly, P.M. Derlet, L.J. Heyderman, and M. Fiebig, Nat. Nanotechnol. **14**, 141 (2019).
[15] J. Nogués and I.K. Schuller, J. Magn. Magn. Mater. **192**, 203 (1999).
[16] S.S.P. Parkin, N. More, and R.K. P., Phys. Rev. Lett. **64**, 2304 (1990).
[17] J.L. Leal and M.H. Kryder, IEEE Trans. Magn. **35**, 800 (1999).
[18] S. Bandiera, R.C. Sousa, Y. Dahmane, C. Ducruet, C. Portemont, V. Baltz, S. Auffret, I.L. Prejbeanu, and B. Dieny, IEEE Magn. Lett. **1**, 3000204 (2010).
[19] J. Chatterjee, S. Auffret, R. Sousa, P. Coelho, I.L. Prejbeanu, and B. Dieny, Sci. Rep. **8**, 11724 (2018).
[20] I. Dzyaloshinsky, J. Phys. Chem. Solids **4**, 241 (1958).
[21] T. Moriya, Phys. Rev. **120**, 91 (1960).
[22] K. Xia, W. Zhang, M. Lu, and H. Zhai, Phys. Rev. B **55**, 12561 (1997).
[23] A. Crépieux and C. Lacroix, J. Magn. Magn. Mater. **182**, 341 (1998).
[24] A.N. Bogdanov and U.B. Rößler, Phys. Rev. Lett. **87**, 37203 (2001).
[25] U.K. Rößler, A.A. Leonov, and A.N. Bogdanov, J. Phys. Conf. Ser **303**, 12105 (2011).
[26] D.H. Kim, M. Haruta, H.W. Ko, G. Go, H.J. Park, T. Nishimura, D.Y. Kim, T. Okuno, Y. Hirata, Y. Futakawa, H. Yoshikawa, W. Ham, S. Kim, H. Kurata, A. Tsukamoto, Y. Shiota, T. Moriyama, S.B. Choe, K.J. Lee, and T. Ono, Nat. Mater. **18**, 685 (2019).
[27] M. Heide, G. Bihlmayer, and S. Blügel, Phys. Rev. B **78**, 140403 (2008).
[28] A. Belabbes, G. Bihlmayer, F. Bechstedt, S. Blügel, and A. Manchon, Phys. Rev. Lett. **117**, 247202 (2016).
[29] T. Kikuchi, T. Koretsune, R. Arita, and G. Tatara, Phys. Rev. Lett. **116**, 247201 (2016).
[30] H. Jia, B. Zimmermann, G. Michalicek, G. Bihlmayer, and S. Blügel, Phys. Rev. Mater. **4**, 024405 (2020).
[31] A. Belabbes, G. Bihlmayer, S. Blügel, and A. Manchon, Sci. Rep. **6**, 1 (2016).
[32] H. Yang, G. Chen, A.A.C. Cotta, A.T. N'DIaye, S.A. Nikolaev, E.A. Soares, W.A.A. MacEdo, K. Liu, A.K. Schmid, A. Fert, and M. Chshiev, Nat. Mater. **17**, 605 (2018).
[33] S. Vélez, J. Schaab, M.S. Wörnle, M. Müller, E. Gradauskaite, P. Welter, C. Gutgsell, C. Nistor, C.L. Degen, M. Trassin, M. Fiebig, and P. Gambardella, Nat. Commun. **10**, 4750 (2019).
[34] C.O. Avci, E. Rosenberg, L. Caretta, F. Büttner, M. Mann, C. Marcus, D. Bono, C.A. Ross, and G.S.D. Beach, Nat. Nanotechnol. **14**, 561 (2019).
[35] A. Thiaville, S. Rohart, É. Jué, V. Cros, and A. Fert, EPL **100**, 57002 (2012).
[36] K.S. Ryu, L. Thomas, S.H. Yang, and S. Parkin, Nat. Nanotechnol. **8**, 527 (2013).
[37] S. Emori, U. Bauer, S.M. Ahn, E. Martinez, and G.S.D. Beach, Nat. Mater. **12**, 611 (2013).
[38] J.H. Franken, M. Hoeijmakers, R. Lavrijsen, J.T. Kohlhepp, H.J.M. Swagten, B. Koopmans, E. Van Veldhoven, and D.J. Maas, J. Appl. Phys. **109**, 07D504 (2011).
[39] G. Chen, T. Ma, A.T. N'Diaye, H. Kwon, C. Won, Y. Wu, and A.K. Schmid, Nat. Commun. **4**, 2671

(2013).

[40] A. Hrabec, N.A. Porter, A. Wells, M.J. Benitez, G. Burnell, S. McVitie, D. McGrouther, T.A. Moore, and C.H. Marrows, Phys. Rev. B **90**, 020402 (2014).

[41] M. Bode, M. Heide, K. Von Bergmann, P. Ferriani, S. Heinze, G. Bihlmayer, A. Kubetzka, O. Pietzsch, S. Blügel, and R. Wiesendanger, Nature **447**, 190 (2007).

[42] S. Meckler, N. Mikuszeit, A. Preßler, E.Y. Vedmedenko, O. Pietzsch, and R. Wiesendanger, Phys. Rev. Lett. **103**, 157201 (2009).

[43] S. Mühlbauer, B. Binz, F. Jonietz, C. Pfleiderer, A. Rosch, A. Neubauer, R. Georgii, and P. Böni, Science **323**, 915 (2009).

[44] K. Von Bergmann, A. Kubetzka, O. Pietzsch, and R. Wiesendanger, J. Phys. Condens. Matter **26**, 394002 (2014).

[45] B. Dupé, M. Hoffmann, C. Paillard, and S. Heinze, Nat. Commun. **5**, 4030 (2014).

[46] A. Soumyanarayanan, M. Raju, A.L.G. Oyarce, A.K.C. Tan, M.Y. Im, A.P. Petrovic, P. Ho, K.H. Khoo, M. Tran, C.K. Gan, F. Ernult, and C. Panagopoulos, Nat. Mater. **16**, 898 (2017).

[47] M. Steinbrecher, R. Rausch, K.T. That, J. Hermenau, A.A. Khajetoorians, M. Potthoff, R. Wiesendanger, and J. Wiebe, Nat. Commun. **9**, 2853 (2018).

[48] F. Büttner, I. Lemesh, and G.S.D. Beach, Sci. Rep. **8**, 4464 (2018).

[49] A. Bernand-Mantel, L. Camosi, A. Wartelle, N. Rougemaille, M. Darques, and L. Ranno, SciPost Phys **4**, 27 (2018).

[50] E.Y. Vedmedenko, L. Udvardi, P. Weinberger, and R. Wiesendanger, Phys. Rev. B **75**, 104431 (2007).

[51] I. Gross, L.J. Martínez, J.P. Tetienne, T. Hingant, J.F. Roch, K. Garcia, R. Soucaille, J.P. Adam, J. V. Kim, S. Rohart, A. Thiaville, J. Torrejon, M. Hayashi, and V. Jacques, Phys. Rev. B **94**, 064413 (2016).

[52] O. Boulle, J. Vogel, H. Yang, S. Pizzini, D. De Souza Chaves, A. Locatelli, T.O. Menteş, A. Sala, L.D. Buda-Prejbeanu, O. Klein, M. Belmeguenai, Y. Roussigné, A. Stashkevich, S. Mourad Chérif, L. Aballe, M. Foerster, M. Chshiev, S. Auffret, I.M. Miron, and G. Gaudin, Nat. Nanotechnol. **11**, 449 (2016).

[53] S. V. Grigoriev, Y.O. Chetverikov, D. Lott, and A. Schreyer, Phys. Rev. Lett. **100**, 197203 (2008).

[54] S. V. Grigoriev, D. Lott, Y.O. Chetverikov, A.T.D. Grünwald, R.C.C. Ward, and A. Schreyer, Phys. Rev. B **82**, 195432 (2010).

[55] A. Fernández-Pacheco, E. Vedmedenko, F. Ummelen, R. Mansell, D. Petit, and R.P. Cowburn, Nat. Mater. **18**, 679 (2019).

[56] E.Y. Vedmedenko, P. Riego, J.A. Arregi, and A. Berger, Phys. Rev. Lett. **122**, 257202 (2019).

[57] D.S. Han, K. Lee, J.P. Hanke, Y. Mokrousov, K.W. Kim, W. Yoo, Y.L.W. van Hees, T.W. Kim, R. Lavrijsen, C.Y. You, H.J.M. Swagten, M.H. Jung, and M. Kläui, Nat. Mater. **18**, 703 (2019).

[58] A.A. Khajetoorians, M. Steinbrecher, M. Ternes, M. Bouhassoune, M. Dos Santos Dias, S. Lounis, J. Wiebe, and R. Wiesendanger, Nat. Commun. **7**, 10620 (2016).

[59] M. Schmitt, P. Moras, G. Bihlmayer, R. Cotsakis, M. Vogt, J. Kemmer, A. Belabbes, P.M. Sheverdyaeva, A.K. Kundu, C. Carbone, S. Blügel, and M. Bode, Nat. Commun. **10**, 2610 (2019).

[60] Z. Luo, T.P. Dao, A. Hrabec, J. Vijayakumar, A. Kleibert, M. Baumgartner, E. Kirk, J. Cui, T. Savchenko, G. Krishnaswamy, L.J. Heyderman, and P. Gambardella, Science **363**, 1435 (2019).

[61] R.A. Gallardo, T. Schneider, A.K. Chaurasiya, A. Oelschlägel, S.S.P.K. Arekapudi, A. Roldán-Molina, R. Hübner, K. Lenz, A. Barman, J. Fassbender, J. Lindner, O. Hellwig, and P. Landeros, Phys. Rev. Appl. **12**, 34012 (2019).

[62] M. Belmeguenai, J.P. Adam, Y. Roussigné, S. Eimer, T. Devolder, J. Von Kim, S.M. Cherif, A. Stashkevich, and A. Thiaville, Phys. Rev. B **91**, 180405 (2015).

[63] A. Manchon, C. Ducruet, L. Lombard, S. Auffret, B. Rodmacq, B. Dieny, S. Pizzini, J. Vogel, V. Uhlíř, M. Hochstrasser, and G. Panaccione, J. Appl. Phys. **104**, 043914 (2008).

[64] I. Gross, W. Akhtar, V. Garcia, L.J. Martínez, S. Chouaieb, K. Garcia, C. Carrétéro, A. Barthélémy, P. Appel, P. Maletinsky, J. V. Kim, J.Y. Chauleau, N. Jaouen, M. Viret, M. Bibes, S. Fusil, and V. Jacques, Nature **549**, 252 (2017).

[65] M. Morin, E. Canévet, A. Raynaud, M. Bartkowiak, D. Sheptyakov, V. Ban, M. Kenzelmann, E. Pomjakushina, K. Conder, and M. Medarde, Nat. Commun. **7**, 13758 (2016).

[66] W.H. Kleiner, Phys. Rev. **142**, 318 (1966).

[67] L. Šmejkal, R. González-Hernández, T. Jungwirth, and J. Sinova, ArXiv E-Prints 1901.00445 (2019).

[68] J.Y. Chauleau, T. Chirac, S. Fusil, V. Garcia, W. Akhtar, J. Tranchida, P. Thibaudeau, I. Gross, C. Blouzon, A. Finco, M. Bibes, B. Dkhil, D.D. Khalyavin, P. Manuel, V. Jacques, N. Jaouen, and M. Viret, Nat. Mater. **19**, 386 (2019).

[69] A. Manchon, J. Železný, I.M. Miron, T. Jungwirth, J. Sinova, A. Thiaville, K. Garello, and P. Gambardella, Rev. Mod. Phys. **91**, 035004 (2019).

[70] I.M. Miron, K. Garello, G. Gaudin, P.-J. Zermatten, M. V. Costache, S. Auffret, S. Bandiera, B.

Rodmacq, A. Schuhl, and P. Gambardella, Nature **476**, 189 (2011).

[71] L. Liu, C.F. Pai, Y. Li, H.W. Tseng, D.C. Ralph, and R.A. Buhrman, Science **336**, 555 (2012).

[72] K. Garello, I.M. Miron, C.O. Avci, F. Freimuth, Y. Mokrousov, S. Blügel, S. Auffret, O. Boulle, G. Gaudin, and P. Gambardella, Nat. Nanotechnol. **8**, 587 (2013).

[73] S. Fukami, C. Zhang, S. DuttaGupta, A. Kurenkov, and H. Ohno, Nat. Mater. **15**, 535 (2016).

[74] M. Baumgartner, K. Garello, J. Mendil, C.O. Avci, E. Grimaldi, C. Murer, J. Feng, M. Gabureac, C. Stamm, Y. Acremann, S. Finizio, S. Wintz, J. Raabe, and P. Gambardella, Nat. Nanotechnol. **12**, 980 (2017).

[75] S.S.P. Parkin, M. Hayashi, and L. Thomas, Science **320**, 190 (2008).

[76] K. Shigeto, T. Shinjo, and T. Ono, Appl. Phys. Lett. **75**, 2815 (1999).

[77] D. Ravelosona, S. Mangin, Y. Lemaho, J.A. Katine, B.D. Terris, and E.E. Fullerton, Phys. Rev. Lett. **96**, 186604 (2006).

[78] T. Phung, A. Pushp, L. Thomas, C. Rettner, S.H. Yang, K.S. Ryu, J. Baglin, B. Hughes, and S. Parkin, Nano Lett. **15**, 835 (2015).

[79] T.P. Dao, M. Müller, Z. Luo, M. Baumgartner, A. Hrabec, L.J. Heyderman, and P. Gambardella, Nano Lett. **19**, 5930 (2019).

[80] P.P.J. Haazen, E. Murè, J.H. Franken, R. Lavrijsen, H.J.M. Swagten, and B. Koopmans, Nat. Mater. **12**, 299 (2013).

[81] J.H. Franken, M. Herps, H.J.M. Swagten, and B. Koopmans, Sci. Rep. **4**, 5248 (2014).

[82] Z. Luo, A. Hrabec, T.P. Dao, G. Sala, S. Finizio, J. Feng, S. Mayr, J. Raabe, P. Gambardella, and L.J. Heyderman, Nature **579**, 214 (2020).

[83] S. Monso, B. Rodmacq, S. Auffret, G. Casali, F. Fettar, B. Gilles, B. Dieny, and P. Boyer, Appl. Phys. Lett. **80**, 4157 (2002).

[84] B. Dieny and M. Chshiev, Rev. Mod. Phys. **89**, 025008 (2017).

[85] Ł. Frackowiak, P. Kuświk, G.D. Chaves-O'flynn, M. Urbaniak, M. Matczak, P.P. Michałowski, A. Maziewski, M. Reginka, A. Ehresmann, and F. Stobiecki, Phys. Rev. Lett. **124**, 047203 (2020).

[86] T. Devolder, J. Ferré, C. Chappert, H. Bernas, J.P. Jamet, and V. Mathet, Phys. Rev. B **64**, 644151 (2001).

[87] M. Urbánek, L. Flajšman, V. Křiáková, J. Gloss, M. Horký, M. Schmid, and P. Varga, APL Mater. **6**, 060701 (2018).

[88] E. Albisetti, D. Petti, M. Pancaldi, M. Madami, S. Tacchi, J. Curtis, W.P. King, A. Papp, G. Csaba, W. Porod, P. Vavassori, E. Riedo, and R. Bertacco, Nat. Nanotechnol. **11**, 545 (2016).

[89] H. Ohno, D. Chiba, F. Matsukura, T. Omiya, E. Abe, T. Dietl, Y. Ohno, and K. Ohtani, Nature **408**, 944 (2000).

[90] M. Weisheit, S. Fähler, A. Marty, Y. Souche, C. Poinsignon, and D. Givord, Science **315**, 349 (2007).

[91] T. Maruyama, Y. Shiota, T. Nozaki, K. Ohta, N. Toda, M. Mizuguchi, A.A. Tulapurkar, T. Shinjo, M. Shiraishi, S. Mizukami, Y. Ando, and Y. Suzuki, Nat. Nanotechnol. **4**, 158 (2009).

[92] C. Bi, Y. Liu, T. Newhouse-Illige, M. Xu, M. Rosales, J.W. Freeland, O. Mryasov, S. Zhang, S.G.E. Te Velthuis, and W.G. Wang, Phys. Rev. Lett. **113**, 267202 (2014).

[93] U. Bauer, S. Emori, and G.S.D. Beach, Nat. Nanotechnol. **8**, 411 (2013).

[94] M. Schott, A. Bernand-Mantel, L. Ranno, S. Pizzini, J. Vogel, H. Béa, C. Baraduc, S. Auffret, G. Gaudin, and D. Givord, Nano Lett. **17**, 3006 (2017).

[95] I. Shorubalko, L. Pillatsch, and I. Utke, in *Nanosci. Technol.* (Springer Verlag, 2016), pp. 355–393.

[96] T. Srivastava, M. Schott, R. Juge, V. Křižáková, M. Belmeguenai, Y. Roussigné, A. Bernand-Mantel, L. Ranno, S. Pizzini, S.M. Chérif, A. Stashkevich, S. Auffret, O. Boulle, G. Gaudin, M. Chshiev, C. Baraduc, and H. Béa, Nano Lett. **18**, 4871 (2018).

[97] H. Nomura, T. Furuta, K. Tsujimoto, Y. Kuwabiraki, F. Peper, E. Tamura, S. Miwa, M. Goto, R. Nakatani, and Y. Suzuki, Jpn. J. Appl. Phys. **58**, 070901 (2019).

[98] H. Arava, N. Leo, D. Schildknecht, J. Cui, J. Vijayakumar, P.M. Derlet, A. Kleibert, and L.J. Heyderman, Phys. Rev. Appl. **11**, 054086 (2019).

[99] F. Garcia-Sanchez, P. Borys, R. Soucaille, J.P. Adam, R.L. Stamps, and J. Von Kim, Phys. Rev. Lett. **114**, 247206 (2015).

[100] O.M. Volkov, A. Kákay, F. Kronast, I. Mönch, M.A. Mawass, J. Fassbender, and D. Makarov, Phys. Rev. Lett. **123**, 077201 (2019).

[101] V.P. Kravchuk, D.D. Sheka, A. Kákay, O.M. Volkov, U.K. Rößler, J. Van Den Brink, D. Makarov, and Y. Gaididei, Phys. Rev. Lett. **120**, 067201 (2018).

[102] A. Fernández-Pacheco, L. Skoric, J.M. De Teresa, J. Pablo-Navarro, M. Huth, and O. V. Dobrovolskiy, Materials (Basel). **13**, 3774 (2020).

[103] Y. Gaididei, V.P. Kravchuk, and D.D. Sheka, Phys. Rev. Lett. **112**, 257203 (2014).

[104] C. Phatak, Y. Liu, E.B. Gulsoy, D. Schmidt, E. Franke-Schubert, and A. Petford-Long, Nano Lett.

**14**, 759 (2014).

[105] D. Sanz-Hernández, A. Hierro-Rodriguez, C. Donnelly, J. Pablo-Navarro, A. Sorrentino, E. Pereiro, C. Magén, S. McVitie, J.M. de Teresa, S. Ferrer, P. Fischer, and A. Fernández-Pacheco, ACS Nano (2020).

[106] S. Eslami, J.G. Gibbs, Y. Rechkemmer, J. Van Slageren, M. Alarcón-Correa, T.C. Lee, A.G. Mark, G.L.J.A. Rikken, and P. Fischer, ACS Photonics **1**, 1231 (2014).

[107] O.M. Volkov, U.K. Rößler, J. Fassbender, and D. Makarov, J. Phys. D. Appl. Phys. **52**, 345001 (2019).

[108] H. Maurenbrecher, J. Mendil, G. Chatzipirpiridis, M. Mattmann, S. Pané, B.J. Nelson, and P. Gambardella, Appl. Phys. Lett. **112**, 242401 (2018).